\documentclass[doublespacing]{elsart}

\usepackage{graphicx}

\usepackage{amssymb}

\begin{document}

\begin{frontmatter}


\title{$^{85}$Kr and $^{39}${Ar} background in GENIUS}
\author{H.V. Klapdor Kleingrothaus\corauthref{cor1}\thanksref{label1}}
\ead{klapdor@gustav.mpi-hd.mpg.}
\author[label2]{and C.Tomei}
\address{Max-Planck-Institut f\"ur Kernphysik, Postfach
  103980, Heidelberg 69029, Germany}
\thanks[label1]{Spokesman of the GENIUS collaboration.}
\thanks[label2]{On leave of the University of L'Aquila, Italy.}
\corauth[cor1]{Corresponding author. Tel.: +49-6221-516259; fax: +49-6221-516540.}

\begin{abstract}
GENIUS is a proposal for a large scale detector of rare events like double beta decay, cold 
dark matter and low-energy solar neutrinos in real time. The idea of
GENIUS is to operate a large amount of ``naked'' Ge detectors in
liquid nitrogen, with the aim of reducing the background down to a
level of 10$^{-3}$ counts/kg keV y. In this work we investigate the
contribution to the background of GENIUS coming from argon
($^{39}${Ar}) and krypton ($^{85}$Kr) contamination in the liquid nitrogen.  
\end{abstract}

\end{frontmatter}

\section{Introduction}
\label{intro}
GENIUS(GErmanium in liquid Nitrogen Underground
Setup) \cite{ringb,prop_genius} is a proposal 
for a large scale detector of rare events like double beta decay, cold 
dark matter and low-energy solar neutrinos in real time. The idea of
GENIUS is to operate a large amount of ``naked'' Ge detectors in
liquid nitrogen, which has been shown to work already in
\cite{ringb}.\\ 
The proposed scale of the GENIUS experiment is a nitrogen tank of about 
12\,m diameter and 12\,m height with 100\,kg of natural Ge or 100\,kg of 
enriched $^{76}$Ge, in the dark matter and double beta decay versions, 
respectively, suspended in its center. 
With this configuration, which removes most of the material in the vicinity of 
the detectors, the background can be strongly reduced with respect to 
conventionally operated detectors. Moreover, the liquid nitrogen acts at the 
same time as cooling medium and as a shield against external radioactivity.\\
It is clear that residual impurities in the liquid nitrogen, like 
$^{238}$U and $^{232}$Th contaminations, $^{222}$Rn, $^{40}$K and $^{60}$Co, 
can be a source of background for the GENIUS experiment and for this reason 
their contribution has to be deeply investigated. 
This has been done in \cite{ringb,prop_genius,nim_genius,eur_genius}, where we
show, for example, that the contribution of the mentioned isotopes to
the background of GENIUS, in the low-energy region, is of
the order of $3 \cdot 10^{-4}$ counts/kg keV y 
for $^{222}$Rn and $4 \cdot 10^{-4}$ counts/kg keV y for
the remaining isotopes. For these calculations, values of
liquid nitrogen purity measured by the BOREXINO collaborations were
used \cite{Borexino}. For the BOREXINO \cite{Borexino2} experiment the nitrogen purity is a very
crucial point; they realized only recently \cite{HLLG} that they have to address
the problem of argon and krypton background at low energies more seriously. This
is true as well for the KAMLAND \cite{Kamland} experiment, as long as
they aim to measure solar neutrinos, after the first successful
reactor neutrino phase.\\
This motivated us to study in detail the effect of argon and krypton
on the GENIUS background. In this note we investigate the background
contribution coming from the two long-lived 
isotopes $^{85}$Kr (T$_{1/2}$=10.76 y) and $^{39}${Ar} (T$_{1/2}$=269
y) and we show that the potential of GENIUS to look for Dark Matter at low
energies is not affected, while, for solar neutrinos, a reduction of a
factor 10 of the krypton contamination with respect to best-quality
commercially available nitrogen is needed (from which a factor 8 has
already been achieved, see below).

\section{$^{85}$Kr and $^{39}${Ar} simulation}
\label{simul}
\begin{figure}[ht]
\begin{center}
\includegraphics[width=7cm]{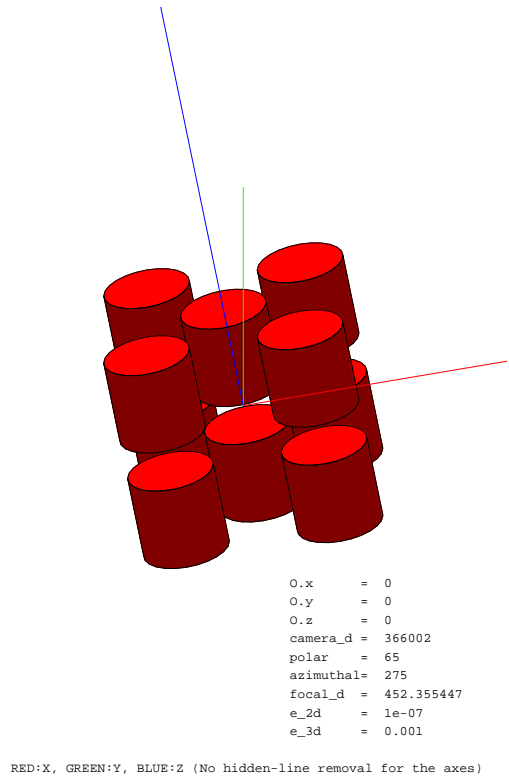}
\end{center}
\caption{Simulated geometry for the GENIUS detectors.}
\label{fig:simul}
\end{figure}

The two isotopes $^{85}$Kr and $^{39}${Ar} decay both via $\beta^-$ decay with 
end point energy of 565.0 keV and 687.1 keV respectively; in addition $^{85}$Kr emits 
four gammas, with energies and intensities listed in Table \ref{tab1}.\\
Both the electrons coming from the $\beta$ decay and the photons from the $\gamma$-lines 
can be a source of background at low energies ($<$500 keV). Since the mean free path of 
the electrons is expected to be very small, only the electrons produced very close to the 
detectors\footnote{Remember that in the GENIUS experiment the detectors are in direct 
contact with liquid nitrogen.} can reach them. In the case of photons, even if produced 
far from the detectors, we expect them to be slowed down and to produce a diffuse background 
at low energies. \\
We calculated the amount of this background by performing a simulation of the GENIUS setup with the 
GEANT4 simulation tool. The configuration of the detectors used for
the simulation is shown in
Fig. \ref{fig:simul}: we assume 10 Ge detectors,
each one with a radius of 6.7 cm and 13.4 cm height, arranged in two layers. The total mass is 
100.534 kg and reproduces the mass of natural germanium foreseen for
the application of GENIUS to Dark Matter search. 
The detectors are suspended in the middle of a cylindrical tank of 12 m diameter and 12 m height. 
According to the measurements performed by the Heidelberg group of the BOREXINO
collaboration \cite{HLLG}, the purest nitrogen available on the market 
has a contamination of 0.03 - 0.2 ppm for argon and about 10
ppt for krypton. 
In the following we will assume a contamination of 0.2 ppm (2.4 $\mu$Bq/m$^3$) from argon
and about 10 ppt (10 $\mu$Bq/m$^3$) from krypton. These values refer to
gaseous nitrogen and have to be multiplied by a factor 650 when
considering liquid nitrogen.
\subsection{Background from $^{39}${Ar}}
\begin{figure}[ht]
\begin{center}
\includegraphics[width=6.5cm]{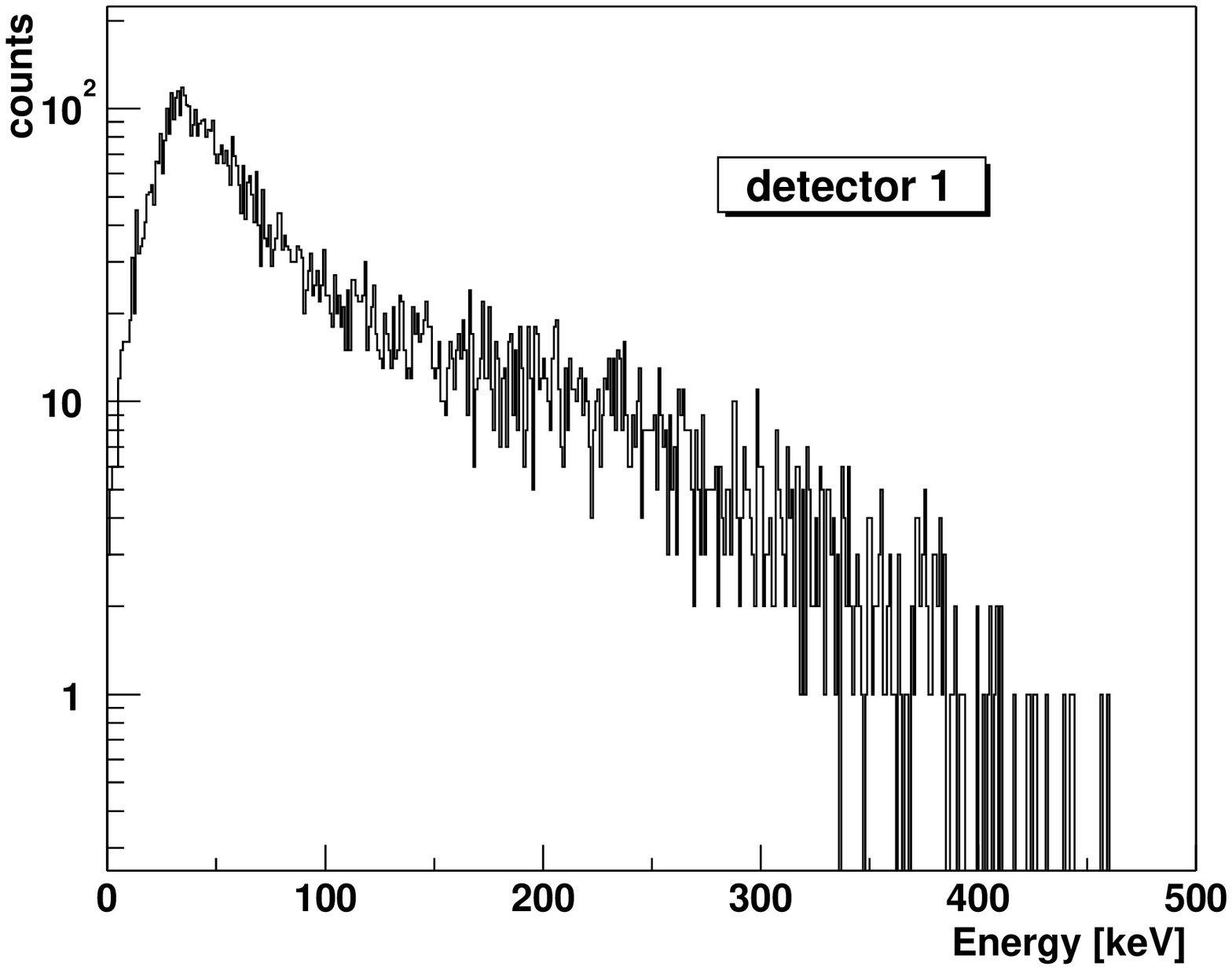}
\includegraphics[width=6.5cm]{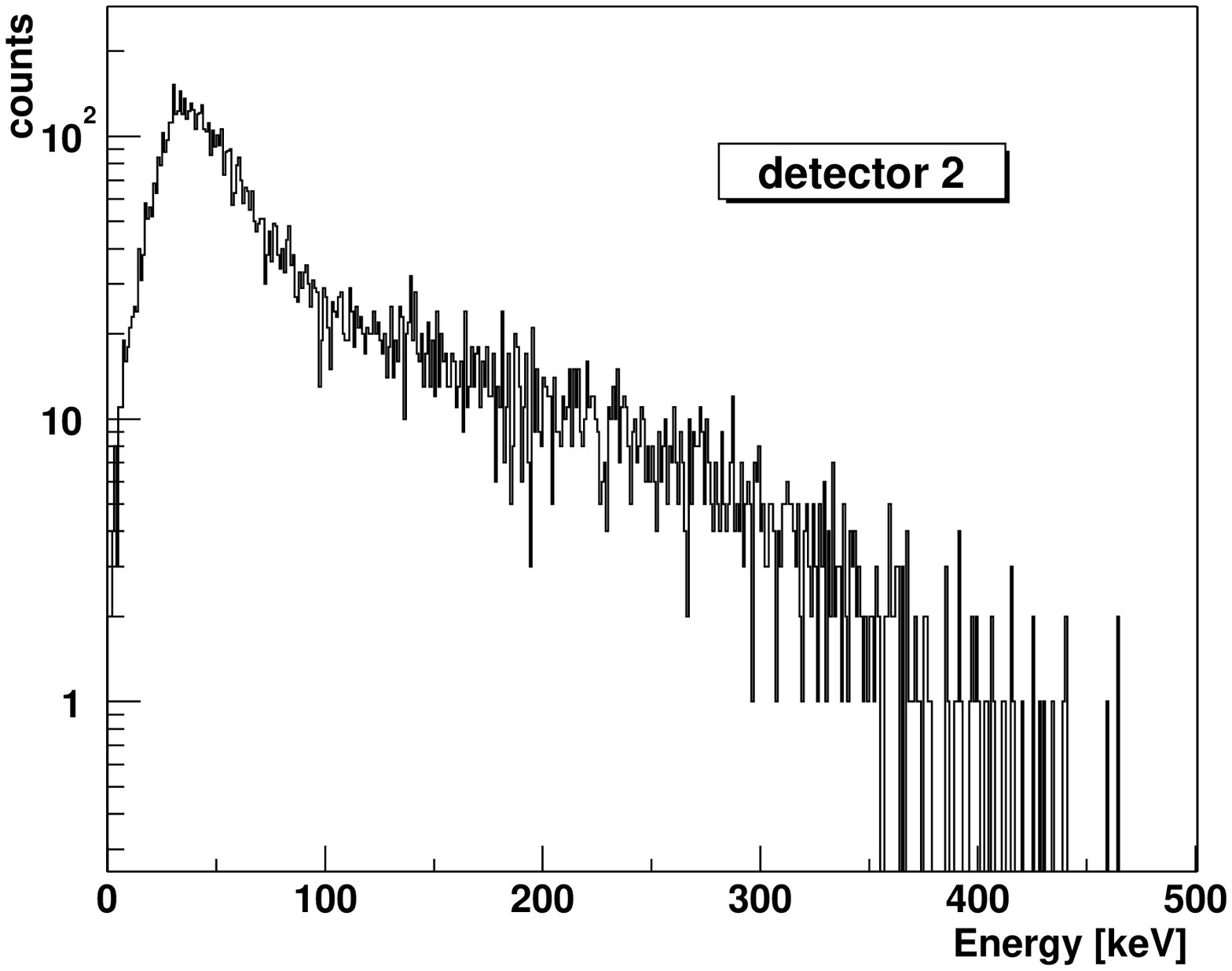}
\end{center}
\caption{Output of the $^{39}${Ar} simulation for 2 of the 10
  detectors in Fig. \ref{fig:simul} (see text).}
\label{fig:argon}
\end{figure}

\begin{figure}[ht]
\begin{center}
\includegraphics[width=7cm,height=5.5cm]{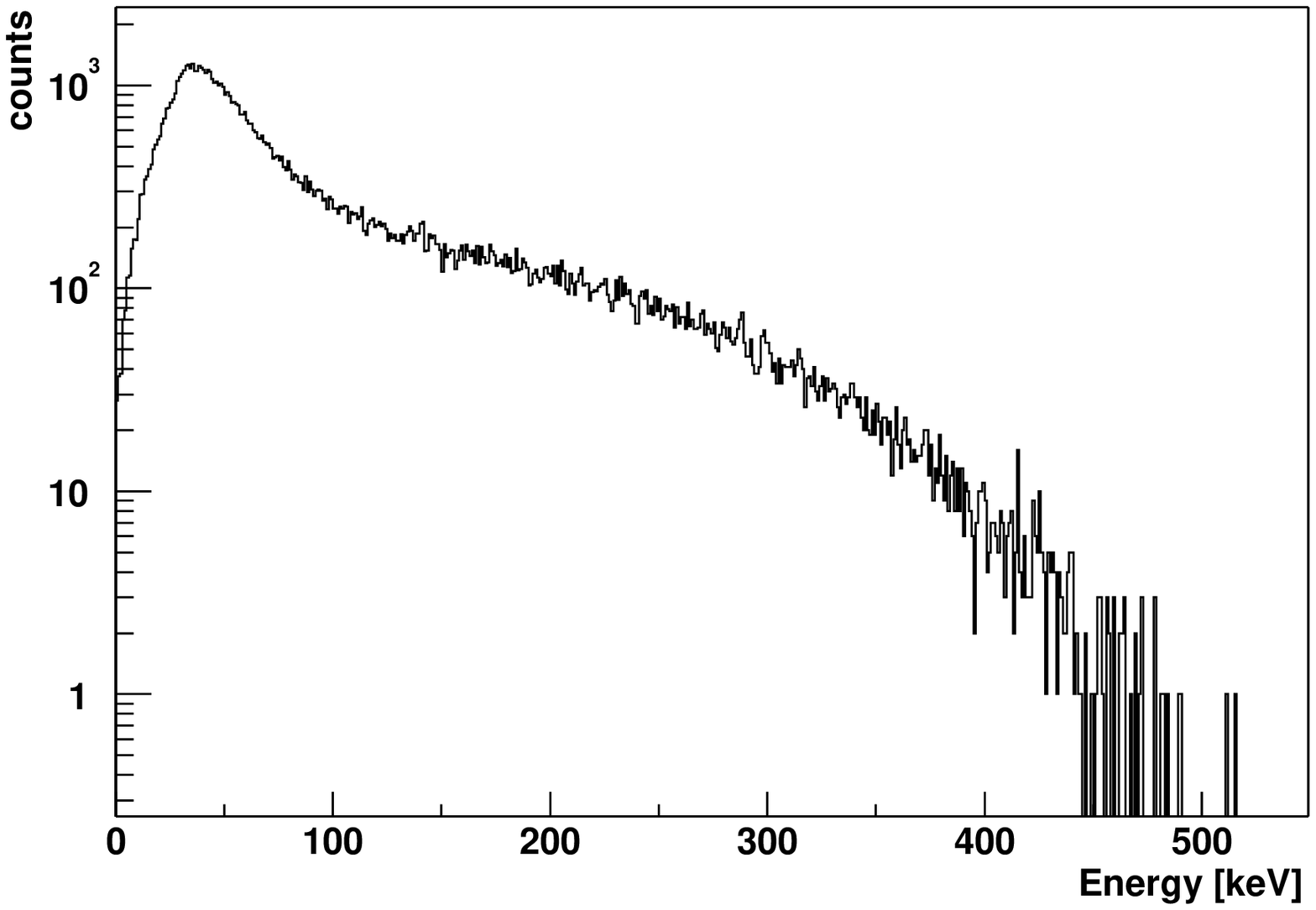}
\includegraphics[width=6.5cm,height=5.5cm]{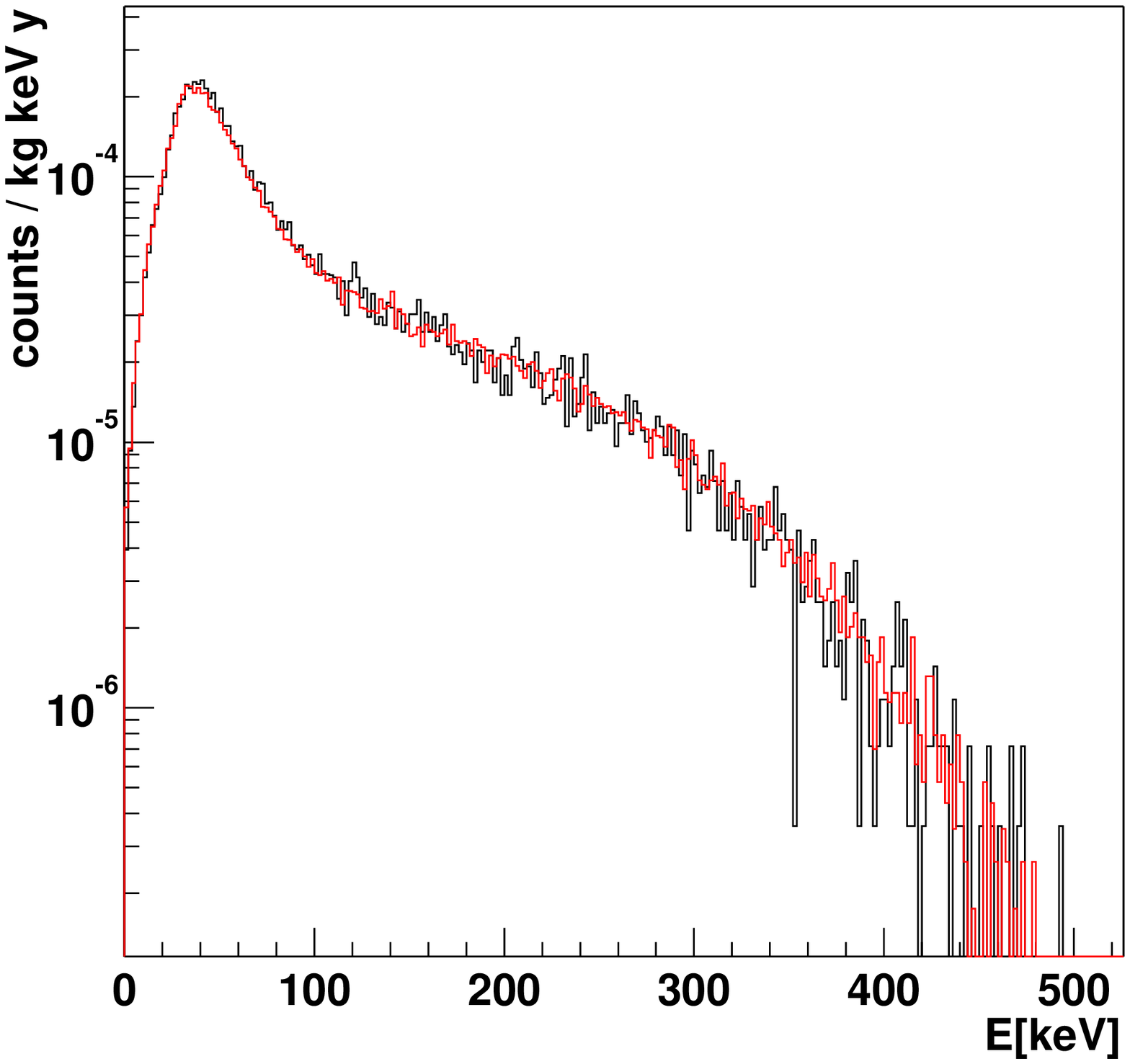}
\end{center}
\caption{$^{39}${Ar} simulation: sum spectrum of all 10 detectors (left) and the same spectrum in 
units of counts/kg keV y (right). The dashed histogram represents the expected $^{39}${Ar} background 
for a volume of liquid nitrogen 8 times bigger (assuming the same argon contamination).}
\label{fig:argon_sum}
\end{figure}

As a first step we simulated the background coming from $^{39}${Ar}. We started assuming that only 
a cylindrical region of 1 m diameter and 1 m height around the center of the tank is filled with 
liquid nitrogen, that means that each detector is surrounded by at least 35 cm of liquid nitrogen 
in each direction. The total number of simulated events is $2.5 \cdot 10^{8}$.
In Fig. \ref{fig:argon} we see the output of this first simulation for two of the detectors 
(the central one and one of the external ones of the upper layer),
Fig. \ref{fig:argon_sum} (left) shows the sum spectrum of all 10 detectors.\\
We can express the background in terms of the usual units (counts/kg
keV y): the result is shown in Fig. \ref{fig:argon_sum} (right).
Increasing the amount of liquid nitrogen around the detectors will not change the number of detected
counts. This is because the electrons coming from $^{39}$Ar-decay have a very short range and so 
only those produced very close to the detectors can contribute to the background. To verify this, 
we performed another simulation, this time assuming that a cylindrical region of 2 m diameter and 
2 m height around the center of the tank is filled with liquid nitrogen (a volume of liquid 8 times
bigger than the previous case); the total number of simulated events was $5 \cdot 10^{8}$.
As we see again from Fig. \ref{fig:argon_sum}, the resulting background
does not change.
\subsection{Background from $^{85}$Kr}
\begin{figure}[ht]
\begin{center}
\includegraphics[width=9cm,height=8cm]{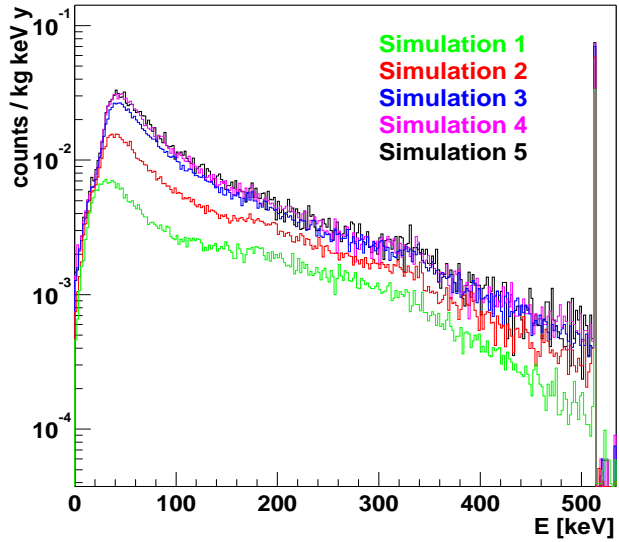}
\end{center}
\caption{Compared results of the $^{85}$Kr simulations described in Table \ref{tab:simul}. Each 
of the five histograms corresponds to a different step of the
simulation and then to a different volume of liquid nitrogen around
the detectors (see Table \ref{tab:simul}).}
\label{fig3}
\end{figure}

\begin{figure}[ht]
\begin{center}
\includegraphics[width=9cm,height=8cm]{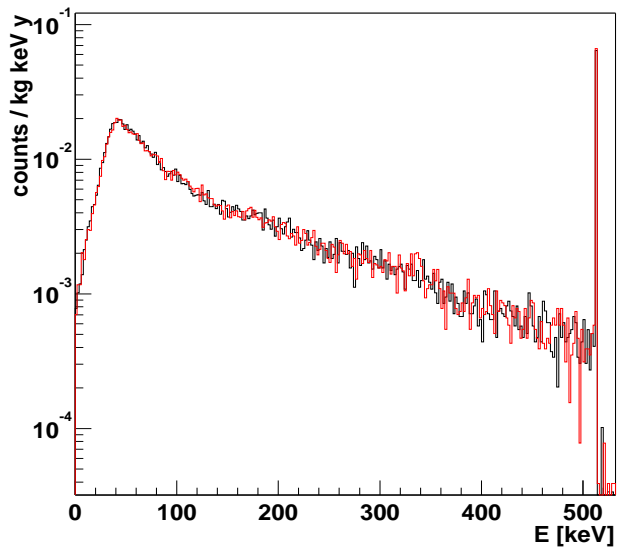}
\end{center}
\caption{$^{85}$Kr background simulated with the geometry of Fig. \ref{fig:simul}, from the first 
simulation (dashed histogram) to the second (solid histogram) we increased the volume of liquid nitrogen,
keeping the same $^{85}$Kr concentration.}
\label{krypton_norm}
\end{figure}

\begin{figure}[ht]
\begin{center}
\includegraphics[width=9cm,height=8cm]{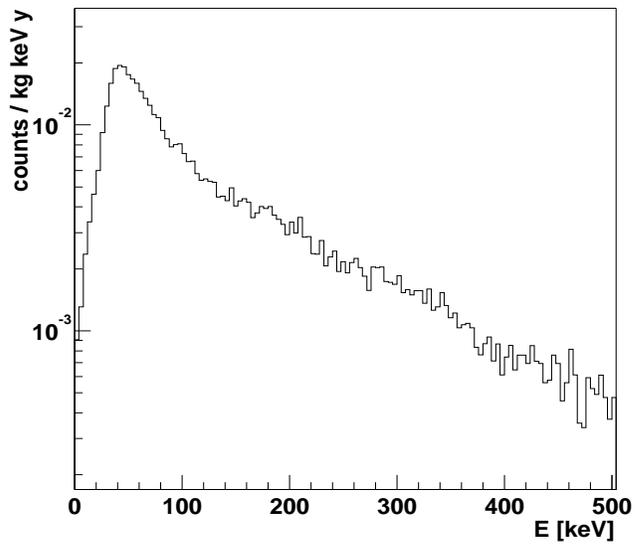}
\end{center}
\caption{Total background for GENIUS in the energy region (0-500) keV,
  coming from the contribution of $^{85}$Kr and $^{39}$Ar.}
\label{total}
\end{figure}

In the case of $^{85}$Kr, a contribution to the background will come both from electrons and from 
the 514 keV photons; photons can travel in liquid nitrogen much longer than electrons and they can 
reach the detectors even if produced far from them.\\
We can have an idea of how far they can travel calculating the mean
free path of photons, with the appropriate energy, in liquid nitrogen.
The total mass attenuation coefficient in nitrogen for $\gamma$-rays with energy of
500 keV is \cite{Tsoulf}:
\begin{equation}
\mu = 8.71 \cdot 10^{-2}\;\;{\rm{cm}}^2/{\rm{g}}.
\end{equation}
Multiplying by the density of liquid nitrogen (0.808 g/cm$^3$) we obtain:
\begin{eqnarray}
\mu &=& 7.04 \cdot 10^{-2}\;\;{\rm{cm}}^{-1} \\
\lambda &=& \mu^{-1} = 14.2\;\;{\rm{cm}} \nonumber
\end{eqnarray}
where $\lambda$ is the mean free path of a 500 keV-photon in liquid nitrogen.\\
From a previous simulation performed with a simpler geometry we obtained some results on the
propagation of the photons coming from $^{85}$Kr. We considered only four detectors in a single 
layer, for a total mass of 13.4 kg, suspended in the middle of a cubic
copper box, filled with liquid nitrogen. This simulation has been
performed in five steps, the details of each step are described in Table \ref{tab:simul}.
The number of simulated events was increased at each step in order to have approximately the same 
statistics in all five cases. In Fig. \ref{fig3} we compare the result of the five simulations 
(after converting each sum spectrum to the usual units): as we can see, increasing the dimensions 
of the copper box the number of decaying atoms increases and,
consequently, we get a higher count rate in the detectors. This
behaviour goes on until we reach a situation were, even adding more liquid 
nitrogen to the setup (and therefore more impurities) we do not see an increase in the detected 
backround: this is due to the fact that the photons coming from the
decay of $^{85}$Kr atoms are 
absorbed by the liquid nitrogen itself.
According to Fig. \ref{fig3}, 100 cm of liquid nitrogen are enough to
absorb the photons coming from the rest of the liquid. \\
We can use this information for a simulation with the more realistic geometry described before (see 
Fig. \ref{fig:simul}). We simulated 1.5 $\cdot 10^{9}$ $^{85}$Kr decays
in a cylindrical volume with a radius of 2 m and 2m height around the
detectors, to be sure that they are surrounded by at least 150 cm of
liquid nitrogen. We obtained the results shown in Fig. \ref{krypton_norm}, which represents 
the total background expected in GENIUS from the decay of $^{85}$Kr. \\
We started also another simulation in a smaller volume (1.5 m radius and 3 m height) and we show in 
the same Fig. \ref{krypton_norm} that the resulting background (red
histogram) is the same, apart from statistical fluctuations. \\
Finally, Fig. \ref{total} shows the total background coming from the contribution of both krypton and 
argon in GENIUS. 
The results on the background indexes are summarized in Table \ref{tab:results}.\\
The final goal of GENIUS is to obtain a level of background of the
order of 10$^{-3}$ counts/kg keV y in the low-energy region. This low
background is requested for real-time detection of solar pp
neutrinos \cite{LowNu2}. Already with 
a background 10 times higher, GENIUS would obtain the planned results in the 
field of Dark Matter searches. It would be able to test the WIMP-parameter space
down to cross sections of the order of 10$^{-9}$pb
\cite{ringb,HVKK}. \\
As shown in Table \ref{tab:results}, the argon contribution to the background will remain
well under the level of 10$^{-3}$ counts/kg keV y, while the
background coming from krypton is expected to be of the level of
10$^{-2}$ counts/kg keV y, which is still acceptable for Dark
Matter detection. A further reduction of the krypton contamination by
a factor of 10 will allow GENIUS to fulfill its background requirement 
also for solar neutrino detection: it must be noted that a krypton purification of a factor 8 has already 
been reached from the Heidelberg group of the Borexino Collaboration
\cite{HLLG}.\\
The GENIUS experiment will also be equipped with a
nitrogen recycling device (through condensation) \cite{prop_genius}, in 
order to avoid accumulation of argon and krypton inside the liquid:
this would happen if the evaporated liquid would be replaced by fresh
nitrogen.

\section{Background considerations for GENIUS-TF}
A test facility for the GENIUS experiment, GENIUS-TF \cite{GENIUS-TF}, is under
construction at the Gran Sasso National Laboratory, with at present six 
natural Ge detectors (about 15 kg) operated in a 50 cm $\times$ 50 cm $\times$ 50 cm
liquid nitrogen container.\\
The goal of GENIUS-TF is to reach a background of 2 - 4 counts/kg keV
y in the low-energy region. The estimated contribution of argon and
krypton contamination to the GENIUS-TF background at low energies,
under the same assumptions as made for GENIUS, is less than 
10$^{-2}$ counts/kg keV y . Even a contamination 100 times higher
would not affect the background in a noticeable way.\\
The problem of the accumulation of argon and krypton in the liquid
nitrogen because of the periodic refilling procedure has been
considered. Assuming that argon and krypton do not evaporate and that we fill the
container when the liquid nitrogen is half of the total volume, we
would have to refill 200 times in order to increase  
the contamination by a factor 100. 

\section{Conclusions}
In the GENIUS detector, a big amount of liquid nitrogen will be used
both as cooling medium and shielding
material and it will be in direct contact with the detectors
\cite{ringb,prop_genius}. For this reason the purity of the
liquid nitrogen not only in terms of radon, but also of argon and
krypton contamination, is a very crucial point. 
The same, and even much more seriously, will be true for the Borexino
experiment, which foresees the use
of gaseous nitrogen to purify the liquid scintillator, and for the KAMLAND experiment when
coming to the solar neutrino detection. For the Borexino experiment,
only a level of nitrogen contamination of 0.36 ppm for argon and 0.16
ppt is acceptable \cite{HLLG}, while the best-quality nitrogen commercially
available has typically 0.03 - 0.2 ppm of argon and about 10 ppt of
krypton. A reduction on the krypton contamination by at least a factor 60 is
then needed.\\
In this work we have shown that the above mentioned commercially
available purity of 0.03 - 0.2 ppm of argon and 10 ppt of krypton would allow the GENIUS
experiment to look for Dark Matter at low energies ($<$ 100 keV), with
a background of the order of 10$^{-2}$ counts/kg keV y, which
corresponds to the planned sensitivity. A further
reduction of a factor 10 (of which a factor of 8 has been already
obtained, see \cite{HLLG}) will bring the background to the level of
10$^{-3}$ counts/kg keV y, which is the requirement of GENIUS for its
use as a solar neutrino detector.

\section*{Acknowledgments}
One of us (C.T.) thanks the Max Planck Institut f\"ur Kernphysik for
the ospitality during her stay in Heidelberg and would like to thank 
Dr. G.Zuzel and B.Freudiger for useful discussions.

\newpage

\begin{table}
\label{tab1}\caption{Gammas from $^{85}$Kr (see Table of Isotopes \cite{TOI}).}
\begin{center}
\begin{tabular}{c|c}
\hline 
E (keV) & I (\%) \\  
\hline
129.820 (12)  &  $<$0.00000043 \\
151.159 (6)   &  0.0000022 (13) \\
362.81 (4)    &  0.0000022 (4)  \\
514.0067 (19) &  0.43 \\
\hline
\end{tabular}
\end{center}
\end{table}

\begin{table}
\label{tab:simul}\caption{Details of the five steps of the $^{85}$Kr simulation.}
\begin{center}
\begin{tabular}{c|c|c|c}
\hline 
Simulation & \# of events & box dimension $[\mathrm{cm}]$ & LiN volume $[\mathrm{cm}^3]$\\  
\hline
1  &  1   $\times$ 10$^7$ &  25 cm &    13080 \\
2  &  9   $\times$ 10$^7$ &  50 cm &   122455 \\
3  &  5   $\times$ 10$^8$ & 100 cm &   997455 \\
4  &  2   $\times$ 10$^9$ & 200 cm &  7997455 \\
5  &  3.5 $\times$ 10$^9$ & 300 cm & 26997455 \\
\hline
\end{tabular}
\end{center}
\end{table}

\newpage

\begin{table}
\label{tab:results}
\caption{Expected background from $^{85}$Kr and $^{39}$Ar in GENIUS in the low-energy region.}
\begin{center}
\begin{tabular}{c|c|c}
\hline 
Element & Energy $[$keV$]$ & Background $[$ counts / kg keV y $]$ \\  
\hline
$^{85}$Kr & 0 - 100 & 1 $\cdot$ 10$^{-2}$  \\
$^{85}$Kr & 0 - 500 & 3.8 $\cdot$ 10$^{-3}$  \\
$^{85}$Kr & 514     &  0.13  \\
$^{39}$Ar & 0 - 100 & 1.1 $\cdot$ 10$^{-4}$  \\
$^{39}$Ar & 0 - 500 & 3.2 $\cdot$ 10$^{-5}$  \\
\hline
\end{tabular}
\end{center}
\end{table}

\end{document}